\documentclass[aps,pra,showpacs,twocolumn,superscriptaddress]{revtex4}
\usepackage{graphicx,latexsym,amssymb}
\usepackage{amsmath}
\usepackage{amsfonts}
\usepackage{times}
\usepackage[colorlinks={true}]{hyperref}
\hypersetup{citecolor={blue}, filecolor={blue}, linkcolor={blue}, urlcolor={blue}}

\begin{document}

\title{Controlling entropic uncertainty bound through memory effects}

\author{G\"{o}ktu\u{g} Karpat}
\email{goktug.karpat@utu.fi}
\affiliation{Turku Center for Quantum Physics, Department of Physics and Astronomy, University of Turku, FI-20014, Turun yliopisto, Finland}
\affiliation{Faculdade de Ci\^encias, UNESP - Universidade Estadual Paulista, Bauru, SP, 17033-360, Brazil}

\author{Jyrki Piilo}
\affiliation{Turku Center for Quantum Physics, Department of Physics and Astronomy, University of Turku, FI-20014, Turun yliopisto, Finland}

\author{Sabrina Maniscalco}
\affiliation{Turku Center for Quantum Physics, Department of Physics and Astronomy, University of Turku, FI-20014, Turun yliopisto, Finland}


\begin{abstract}
One of the defining traits of quantum mechanics is the uncertainty principle which was originally expressed in terms of the standard deviation of two observables. Alternatively, it can be formulated using entropic measures, and can also be generalized by including a memory particle that is entangled with the particle to be measured. Here we consider a realistic scenario where the memory particle is an open system interacting with an external environment. Through the relation of conditional entropy to mutual information, we provide a link between memory effects and the rate of change of conditional entropy controlling the lower bound of the entropic uncertainty relation. Our treatment reveals that the memory effects stemming from the non-Markovian nature of quantum dynamical maps directly control the lower bound of the entropic uncertainty relation in a general way, independently of the specific type of interaction between the memory particle and its environment.
\end{abstract}

\maketitle

\textbf{Introduction}-- Uncertainty principle is a pillar of quantum theory, embodying one of its characteristic traits: inevitable uncertainty limiting our ability to predict the measurement results of two incompatible observables simultaneously. Based on the ideas of Heisenberg \cite{hei27} related to the uncertainty of position $x$ and momentum $p_x$, Kennard \cite{ken27} formulated the first uncertainty relation in terms of the product of standard deviations, i.e., $\Delta x \Delta p_x \geq \hbar/2$. Later, this relation was generalized by Robertson \cite{robert29} for two arbitrary observables $Y$ and $Z$ as $\Delta Y \Delta Z \geq \frac{1}{2}|\langle\psi|[Y,Z]|\psi\rangle|.$ Nevertheless, there are several drawbacks in quantifying uncertainty via standard deviation \cite{entrobook,deutsch83}. Moreover, the uncertainty bound above is state dependent, and it can become trivial when a state $|\psi\rangle$ has zero expectation value for the commutator $[Y,Z]$.

An alternative method is to quantify the uncertainty about the probability distribution for measurement outcomes based on the use of entropic measures \cite{wehner10}. Such an approach is especially meaningful when we are interested in the uncertainty related to the lack of knowledge of possible measurement outcomes. One of the most well known entropic uncertainty relations was proved by Maasen and Uffink \cite{maasen88},
\begin{equation} \label{nomemunc}
H(Q)+H(R)\geq \log_2 \frac{1}{c},
\end{equation}
where the Shannon entropy $H(X)=-\sum_x p(x) \log_2 p(x)$ quantifies the amount of uncertainty about the observable $X\in(Q,R)$ before the result of its measurement is revealed. Here, the probability of the outcome $x$ is denoted by $p(x)$ when a density operator $\rho$ is measured in $X$-basis. Complementarity of the observables $Q$ and $R$ is given by $1/c=1/\max_{i,j}|\langle\psi_i|\phi_j\rangle|^2$, where $|\psi_i\rangle$ and $|\phi_j\rangle$ are the eigenstates of the Hermitian observables $Q$ and $R$, respectively.

We now consider a scenario in which Bob has access to an additional particle serving as a quantum memory (particle $B$), which is entangled with the particle held by Alice (particle $A$). Alice performs measurements on her particle as described by $Q$ and $R$. In this setting, Berta et al. showed \cite{berta10} that a more general uncertainty relation holds
\begin{equation} \label{unc}
S(Q|B)+S(R|B)\geq \log_2 \frac{1}{c} + S(A|B),
\end{equation}
where $S(A|B)=S(\rho_{AB})-S(\rho_B)$ is the conditional entropy. While $S(\rho)=\textmd{tr}[\rho \log_2 \rho]$ denotes the von Neumann entropy, $S(X|B)$ with $X\in(Q,R)$ represents the conditional entropies of the post-measurement states $\rho_{XB}=\sum_j(|\psi_j\rangle\langle\psi_j|\otimes \mathbb{I})\rho_{AB}(|\psi_j\rangle\langle\psi_j|\otimes \mathbb{I})$ after the subsystem $A$ is measured in $X$ basis, $\{|\psi_j\rangle\}$ are the eigenstates of the observable $X$, and $\mathbb{I}$ is the identity matrix. The memory-assisted uncertainty relation in Eq. (\ref{unc}) gave rise to applications related to witnessing entanglement and cryptographic security \cite{berta10}. It has also been verified with two experiments \cite{exp1,exp2}.

Realistic quantum systems interact with their surroundings, resulting in the the loss of characteristic features of quantum theory. The effects of this interaction are described within the framework of open quantum systems \cite{openbook}. From the perspective of memory effects, it is conventional to categorize the dynamics of open systems into two groups. While Markovian evolution leads to the absence of memory effects, where the system monotonically loses information to the environment, non-Markovian features might enable the system to recover some part of the information back from the environment, generating memory effects. The characterization of non-Markovianity \cite{rivas10,hou11,breuer09,lu10,luo12,fanchini14,bylicka14,chr14} and the advantages of memory effects \cite{vasile11,laine14,huelga12,chin12,thorwart09,chin13} are an active field of research. All the same, we should underline that our work is not merely another attempt to contribute to discussions about which measure is better than others. Rather, here we point out to a fundamental operational meaning of memory effects in quantum theory.

In this work, we consider a setting where the memory particle $B$ is an open system interacting with an environment $E$ and thus undergoing non-unitary dynamics described by a \textit{t}-parameterised family of quantum channels $\Phi_t$. Let us assume that both the state of the composite system $AB$ and the environment $E$ are initially pure, and keep in mind the relation of the conditional entropy $S(A|B)$ to the mutual information $I(\rho_{AB})$ [see Eq. (\ref{IS})]. We are thereby able to provide a link, via Eq. (\ref{main}), between memory effects, emerging as a consequence of the backflow of information from the environment $E$ to the memory particle $B$, and the rate of change of the conditional entropy $S(A|B)$. We reveal that memory effects directly control the lower bound of uncertainty associated with the observables $Q$ and $R$. Our approach establishes a general connection between the memory effects and the lower bound of the memory-assisted entropic uncertainty relation, in a way that is independent of the type of non-Markovian noise on the memory particle $B$. We demonstrate the implications of our findings by studying Bob's uncertainty, about the measurement results of two observables $Q$ and $R$, along with its lower bound for dephasing and relaxation models.

\textbf{Preliminaries}--It is convenient to think about uncertainty relations with the help of an uncertainty game \cite{berta10}, taking place between Alice and Bob. Firstly, they agree on two observables, $Q$ and $R$. Then, Bob prepares a particle in a quantum state that he desires and sends it to Alice. Finally, Alice measures the particle she received in one of the two agreed bases and tells her choice to Bob, whose task is then to minimize the uncertainty about the measurement outcomes. In fact, in the absence of quantum memory, Eq. (\ref{nomemunc}) restricts Bob's uncertainty about the measurement on Alice's system.

Provided Bob entangles the particle $A$ that he sends to Alice with an additional memory particle $B$ before the game starts, then the memory-assisted entropic uncertainty relation in Eq. (\ref{unc}) bounds his uncertainty about the outcomes of measurements in $Q$ and $R$ bases on Alice's system. Particularly, the left-hand side of Eq. (\ref{unc}) quantifies Bob's total amount of ignorance about Alice's measurement outcomes, given that Bob has access to the memory particle $B$. Also, there appears an additional term on the right-hand side, namely $S(A|B)$, modifying the lower bound of the uncertainty associated to the observables $Q$ and $R$. It is crucial to emphasize that, unlike its classical counterpart, quantum conditional entropy can take negative values, which by itself paves the way to interesting operational applications \cite{horodecki05}. In the extreme case, where $A$ and $B$ are maximally entangled, Bob can indeed correctly predict the measurement outcomes of two incompatible observables with vanishing uncertainty.

We now clarify how we understand the memory effects in open quantum systems. Such memory effects result from the non-Markovian reduced dynamics of the memory particle $B$, caused by its interaction with an external reservoir. However, non-Markovianity is a multi-faceted phenomenon in the quantum domain and there exists no unique way of capturing all different features of memory effects \cite{rivas10,hou11,breuer09,lu10,luo12,fanchini14,bylicka14,chr14,notagree1,notagree2}. We will characterize them through the non-monotonical behavior of the mutual information under local completely positive trace preserving (CPTP) maps \cite{luo12}, since this approach enjoys a physical interpretation in terms of flow of information between the system and its environment \cite{haseli14}.

\begin{figure}[t]
\includegraphics[width=0.42\textwidth]{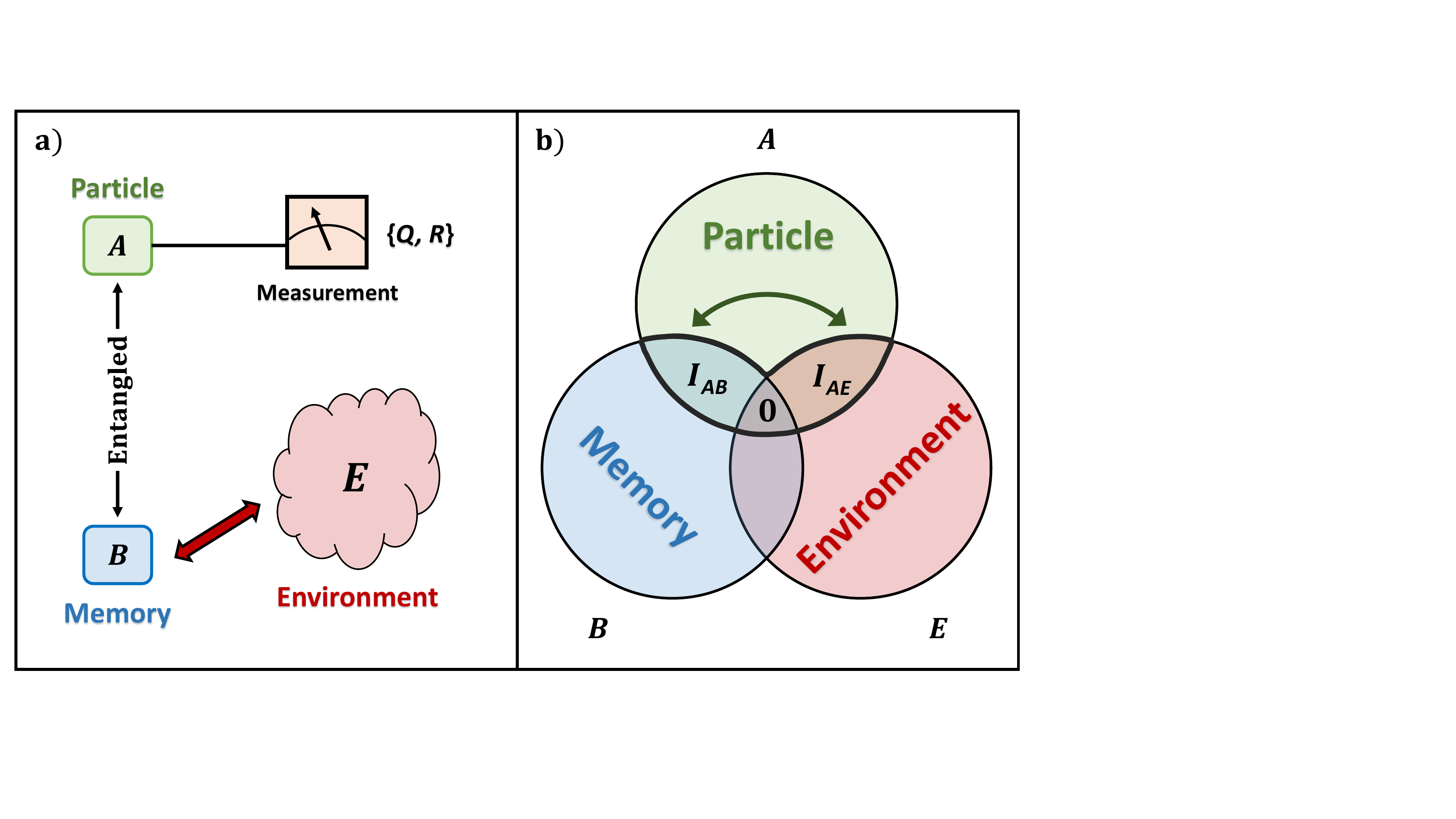}
\caption{a) Schematic representation of our setting where the memory particle $B$ is in interaction with an environment $E$, and the measurements to performed on the particle $A$ are agreed to be in $Q$ and $R$ bases. b) Entropy diagram for the tripartite system $ABE$ showing the information exchange among the parties. The total amount of mutual information enclosed by the thick line, i.e. $I(\rho_{AB})+I(\rho_{AE})$, stays the same throughout the dynamics. Thus, the information that $B$ and $E$ individually share with $A$ flows into each other.}
\label{fig1}
\end{figure}

Quantum mutual information quantifies the total amount correlations in a bipartite state as
\begin{equation}
I(\rho_{AB})=S(\rho_{A})+S(\rho_{B})-S(\rho_{AB}),
\end{equation}
where $\rho_{A}=\textmd{tr}_B[\rho_{AB}]$ and $\rho_{B}=\textmd{tr}_A[\rho_{AB}]$ are density matrices of the reduced systems. Probing the dynamics of mutual information between an open system and an isolated one, Luo, Fu, and Song (LFS) proposed a criterion to identify the memory effects associated with non-Markovian dynamical maps \cite{luo12}. Note that memoryless Markovian maps satisfy the property of divisibility, i.e., $\Phi_t=\Phi_{t,s}\Phi_s$ with $\Phi_{t,s}$ being CPTP and $s \leq t$. If we assume that the map $\Phi_t$ is acting only on the subsystem $B$ and the subsystem $A$ evolves trivially, the absence of memory effects immediately implies
\begin{equation}
I(( \mathbb{I} \otimes \Phi_t)\rho_{AB}) \leq I((\mathbb{I} \otimes \Phi_s)\rho_{AB}),
\end{equation}
at all times $0\leq s \leq t$ for all bipartite states $\rho_{AB}$. Therefore, based on the violation of this inequality, one can detect the presence of memory related to the map $\Phi_t$. That is to say that any revival of $I(\rho_{AB})$ throughout the dynamics is a signature of the memory effects.

\textbf{Main Result}-- In the following, we consider a realistic setting where the memory particle $B$, which is in Bob's possession to improve the uncertainty bound given in Eq. (\ref{unc}), is an open system interacting with an environment $E$, as pictorially sketched in Fig \ref{fig1}a. The motivation for this scenario, where the memory particle $B$ is being affected by the environmental noise, is connected with the fact that the significance of the memory-assisted uncertainty relation in Eq. (\ref{unc}) relies on the quantumness of the memory particle $B$. Therefore, we intend to comprehend the usefulness and relevance of non-Markovian memory effects while the memory particle $B$ becomes classical undergoing a decoherence process.

We first remember that we can write the mutual information $I(\rho_{AB})$ of a bipartite state $\rho_{AB}$ in terms of conditional entropy as $I(\rho_{AB})=S(\rho_A)-S(A|B)$. In our setting, as the bipartite system $AB$ evolves in time, time-dependent mutual information $I(\rho_{AB}(t))$ shared by the particle $A$ to be measured and the memory particle $B$ is given as
\begin{equation} \label{IS}
I(\rho_{AB}(t))=S(\rho_A)-S_{t}(A|B),
\end{equation}
where $S_{t}(A|B)=S(\rho_{AB}(t))-S(\rho_B(t))$. Taking the time derivative of both sides, our main argument simply follows:
\begin{equation} \label{main}
\frac{d}{dt}I(\rho_{AB}(t)) =-\frac{d}{dt}S_{t}(A|B),
\end{equation}
due to the fact that $S(\rho_A)$ is invariant in time. Eq. (\ref{main}) provides a direct connection between the rate of change of the mutual information $I(\rho_{AB}(t))$, whose summation over a certain time interval measures the amount of memory effects, and the rate of change of the conditional entropy $S_{t}(A|B)$, summation of which over the same interval basically controls the uncertainty bound in Eq. (\ref{unc}) since the complementarity $1/c$ is not state dependent. In other words, this relation clearly reveals how the lower bound of the memory-assisted entropic uncertainty relation is reduced via the effects of the memory. In particular, when the memory effects manifest, as signalled by a increase of mutual information $I(\rho_{AB}(t))$, we will observe a decrease in the conditional entropy $S_{t}(A|B)$, which in turn corresponds to a reduction in the lower bound in Eq. (\ref{unc}). At this point, we emphasize that this is a quite interesting and non-trivial result, considering the fact that there exist numerous other quantifiers of non-Markovian memory effects in the literature, none of which can be directly linked to the entropic uncertainty relations as we have demonstrated in this work. In fact, we should also mention that independently of the existence of memory effects in the dynamics, the rate of change of the lower bound of the memory-assisted entropic uncertainty relation is linked to the rate of change of the conditional entropy $S_{t}(A|B)$ and thus to the rate of change of the mutual information $I(\rho_{AB}(t))$. As a result, our approach is not limited to non-Markovian dynamics and it holds for any quantum process defined for the memory particle $B$.

\textbf{Interpretation}--If we assume that both the bipartite system $AB$, and the environment $E$ can be initially described by pure states, then the LFS criterion can be given an information theoretic interpretation in terms of the information exchange between the open system and its environment. There exists a link between the rate of change of the mutual information $I(\rho_{AB}(t))$ shared by the particle $A$ and the memory particle $B$, and the rate of change of the mutual information $I(\rho_{AE}(t))$ between the particle $A$ and the environment $E$ \cite{haseli14},
\begin{equation}
\frac{d}{dt}I(\rho_{AE}(t)) =-\frac{d}{dt}I(\rho_{AB}(t)),
\end{equation}
which follows from the fact that $I(\rho_{AB}(t))+I(\rho_{AE}(t))$ always remains invariant throughout the time evolution. Specifically, if $I(\rho_{AE}(t))$ (which is initially zero) monotonically increases, this will imply a monotonic decrease in $I(\rho_{AB}(t))$. However, in case $I(\rho_{AB}(t))$ rises temporarily, then we will see a reduction in $I(\rho_{AE}(t))$ by the same amount. Note that the ternary mutual information $I(\rho_{ABE})$ vanishes thanks to the pureness of the tripartite system $ABE$. In other words, the memory particle $B$ and the environment $E$ individually exchange the information that they have in common with the particle $A$ back and forth during the dynamics as depicted in Fig. \ref{fig1}b. When the information that $A$ shares with $E$ flows back into the part which $A$ and $B$ have in common, memory effects emerge. Conceptually, this means that the particle $A$ in fact serves as a medium for the correlations hence allowing memory effects to propagate from the environment $E$ to the memory particle $B$. This makes it possible to modify the lower bound of the memory-assisted entropic uncertainty relation. It should be mentioned that even in case of mixed initial environmental states (finite temperature environments), one can directly use the criterion given in terms of the non-monotonic behavior of mutual information between $A$ and $B$, to control the entropic uncertainty bound. Furthermore, our information theoretic interpretation can still be applied with a small modification, where the state of the environment is purified with an additional subsystem $E'$. In this case, our treatment is still fully valid with one difference: the flow of information should now be considered between the subsystems $A$, $B$ and purified environment $EE'$.

\textbf{Examples}--We now elaborate the implications of our result for two types of non-Markovian noise models on the memory particle $B$, which is assumed to be a two-level system. We start to examine our problem for a colored dephasing noise introduced by Daffer et al. in Ref. \cite{daffer04}. Suppose that the dynamics is described by a master equation of the form $\dot{\rho}=K\mathcal{L}\rho$ where $K$ acts on the memory particle $B$ as $K\phi=\int_0^t k(t-t')\phi(t')dt'$, $k(t-t')$ is a kernel function determining the type of memory in the environment, $\rho$ is the density operator of the particle $B$, and $\mathcal{L}$ is a Lindblad superoperator. To study a master equation of this form, we can consider a time-dependent Hamiltonian $H(t)=\hbar\Gamma(t)\sigma_z$, where $\sigma_z$ is the Pauli operator in z-direction and $\Gamma(t)$ is an independent random variable with the statistics of a random telegraph signal. Particularly, it can be expressed as $\Gamma(t)=\alpha n(t)$, where $n(t)$ has a Poisson distribution with a mean equal to $t/2\tau$ and $\alpha$ is a coin-flip random variable having the values $\pm \alpha$. If $\alpha=1$, the dynamics the memory particle $B$ can be described by the Kraus operators
\begin{align}
K_1(\nu) &= \sqrt{[1+\Lambda(\nu)]/2}\mathbb{I}, \\
K_2(\nu) &= \sqrt{[1-\Lambda(\nu)]/2}\sigma_3,
\end{align}
where we have $\Lambda(\nu)=e^{-\nu}[\cos(\mu\nu)+\sin(\mu\nu)/\mu]$, and $\mu=\sqrt{(4\tau)^2-1}$ with $\nu=t/2\tau$ being the scaled time. The parameter $\tau$ controls the degree of non-Markovianity producing the memory effects. In particular, dynamics of $B$ can be obtained using the Kraus operators as
\begin{equation}
\rho_{AB}(\nu) = \sum_{i=1}^{2}K_{i}(\nu)\rho(0)K_{i}^{\dagger}(\nu).
\end{equation}

To study the lower bound of the memory-assisted entropic uncertainty relation, we choose the observables as $Q=\sigma_1$ and $R=\sigma_3$. Also, we set $U(t)\equiv S_{t}(\sigma_1|B)+S_{t}(\sigma_3|B)$ and $UB(t) \equiv \log_2[1/c]+S_{t}(A|B)$, where $U(t)$ and $UB(t)$ stand for uncertainty and uncertainty bound, respectively. Since $\sigma_1$ and $\sigma_3$ are complementary, $\log_2[1/c]$ attains its maximal value, i.e., $\log_2[1/c]=1$. Moreover, the initial states we consider in this work for the bipartite system $AB$ are of the form
\begin{equation}
|\Psi\rangle= \sqrt{a}|00\rangle+\sqrt{b}|10\rangle+\sqrt{c}|11\rangle,
\end{equation}
where the normalization condition holds as $a+b+c=1$.
\begin{figure}[t]
\includegraphics[width=0.47\textwidth]{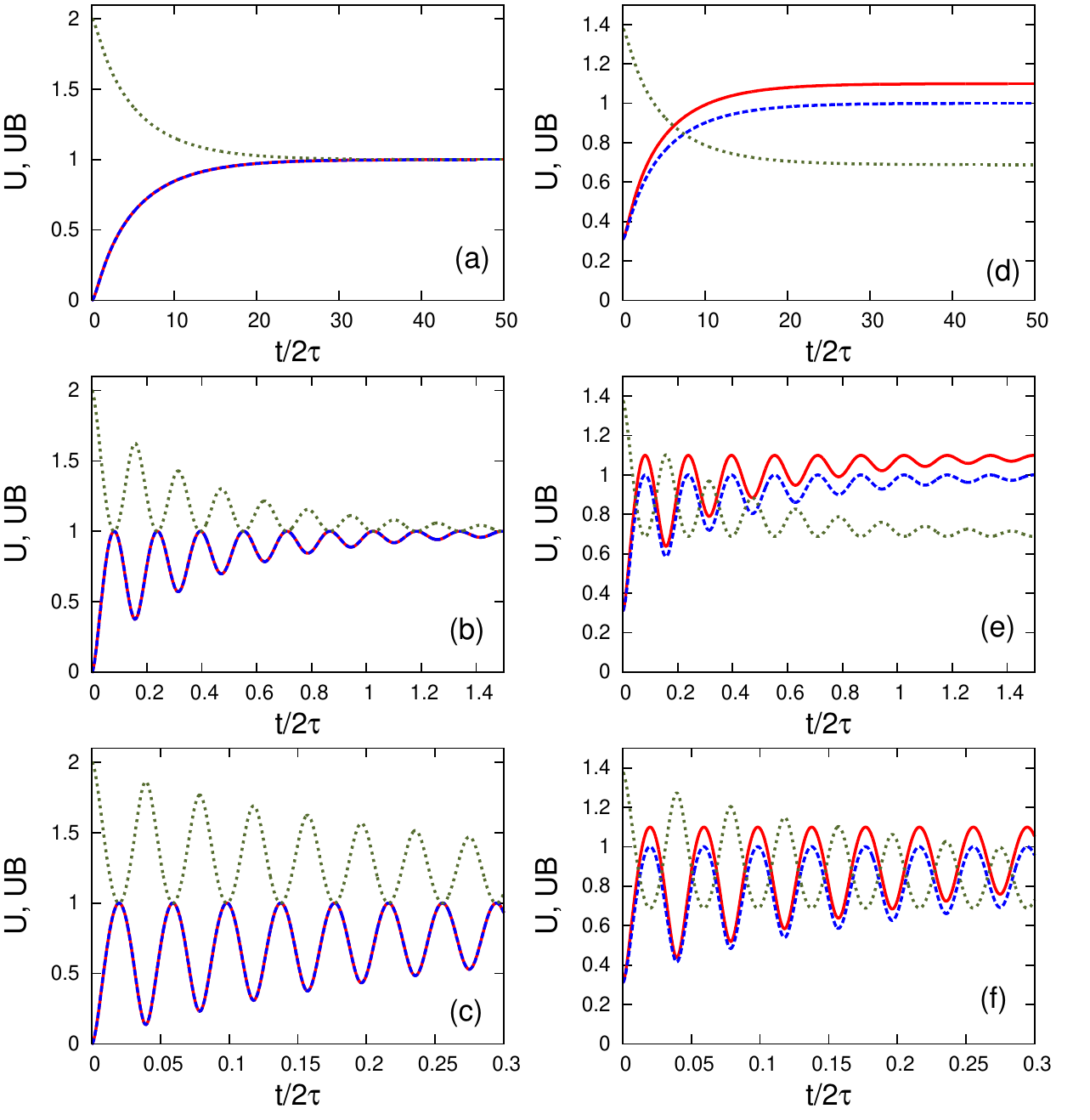}
\caption{Uncertainty $U(t)$ (red solid line), its lower bound $UB(t)$ (blue dashed line) and mutual information $I(\rho_{AB}(t))$ (green dotted line) versus scaled time $t/2\tau$ when the memory particle $B$ is affected by colored dephasing noise. We have $a=0.5$, $b=0$ and $c=0.5$ for the initial state of $AB$ in (a), (b) and (c), and $a=0.5$, $b=0.2$ and $c=0.3$ in (d), (e) and (f). The parameter controlling non-Markovianity is set as $\tau=0.1$ in (a) and (d), $\tau=5$ in (b) and (e), and as $\tau=20$ in (c) and (f). Note that the plots in (a) and (d) shows the Markovian limit of the considered model.}
\label{fig2}
\end{figure}

In Fig. \ref{fig2}, we show the results of our analysis of $U(t)$ (red solid line), $UB(t)$ (blue dashed line), and $I(\rho_{AB}(t))$ (green dotted line) for the dephasing noise. In Fig. \ref{fig2}(a), (b) and (c), we let the initial state of $AB$ to be a maximally entangled state,that is, we set $a=0.5$, $b=0$ and $c=0.5$. While the parameter controlling the degree of non-Markovianity is set to $\tau=0.1$ to show the Markovian limit of the evolution in (a), it is set to $\tau=5$ and $\tau=20$ in (b) and (c), respectively. In the Markovian limit, where memory effects are absent, uncertainty bound $UB(t)$ and uncertainty $U(t)$ can be both observed to be monotonically increasing. However, as a direct consequence of our main result in Eq. (\ref{main}), we see by comparing (b) and (c) that greater amount of non-Markovian memory effects (a greater amount of increase in $I(\rho_{AB}(t))$) implies a greater reduction in $UB(t)$, which is in fact followed by a same amount of reduction in $U(t)$ since the bound is tight in this case. We also note that here the ignorance only comes from the term $S_{t}(\sigma_1|B)$ because $S_{t}(\sigma_3|B)$ vanishes. This example already demonstrates how we can control the lower bound of the memory-assisted uncertainty relation by simply adjusting the degree of memory effects. We emphasize that when the lower bound in Eq. (\ref{unc}) is tight (UB=U), as in this example, memory effects not only control the lower bound but also the actual uncertainty. Moving to the remaining three plots, Fig. \ref{fig2}(d) and (e) and (f), we have $a=0.5$, $b=0.2$ and $c=0.3$ for the initial state of $AB$. Whereas $\tau=0.1$ in Fig. \ref{fig2}(d), we have $\tau=5$ and $\tau=20$ in (e) and (f), respectively. It is clear that we can reach a similar conclusion about the memory effects lowering the bound by comparing these plots. It is important to note that although the bound is not tight for this initial state, the reduction of $UB(t)$ due to the memory effects is followed by a reduction of the actual uncertainty $U(t)$.

Second example deals with a zero temperature relaxation model. The Hamiltonian is
\begin{equation}
H=\omega_0\sigma_{+}\sigma_{-}+\sum_{k}\omega_k a_k^\dagger a_k + (\sigma_{+}B + \sigma_{-}B^\dagger),
\end{equation}
where $\sigma_{\pm}$ represent the operators of the memory $B$ with the transition frequency $\omega_0$, and $B=\sum_k g_k a_k$. The annihilation and creation operators of $E$ are denoted by $a_k$ and $a_k^\dagger$, respectively, having frequencies $\omega_k$. Supposing that the environment has an effective spectral density $J(\omega)= \gamma_0 \lambda^2 / 2\pi[(\omega_0 - \omega)^2 + \lambda^2]$, the dynamics of $B$ is described by the Kraus operators
\begin{align}
M_1(t) &= \begin{pmatrix} 1 & 0\\ 0 & \sqrt{p(t)} \end{pmatrix}, &
M_2(t) &= \begin{pmatrix} 0 & \sqrt{1-p(t)}\\ 0 & 0 \end{pmatrix},
\end{align}
where $p(t)=e^{-\lambda t} [ \cos{\left(dt/2\right)} +(\lambda/d)\sin{\left(dt/2\right)} ]^2$. Here $d=\sqrt{2\gamma_0\lambda-\lambda^2}$ and $\lambda/\gamma_0$ controls non-Markovianity.

\begin{figure}[t]
\includegraphics[width=0.47\textwidth]{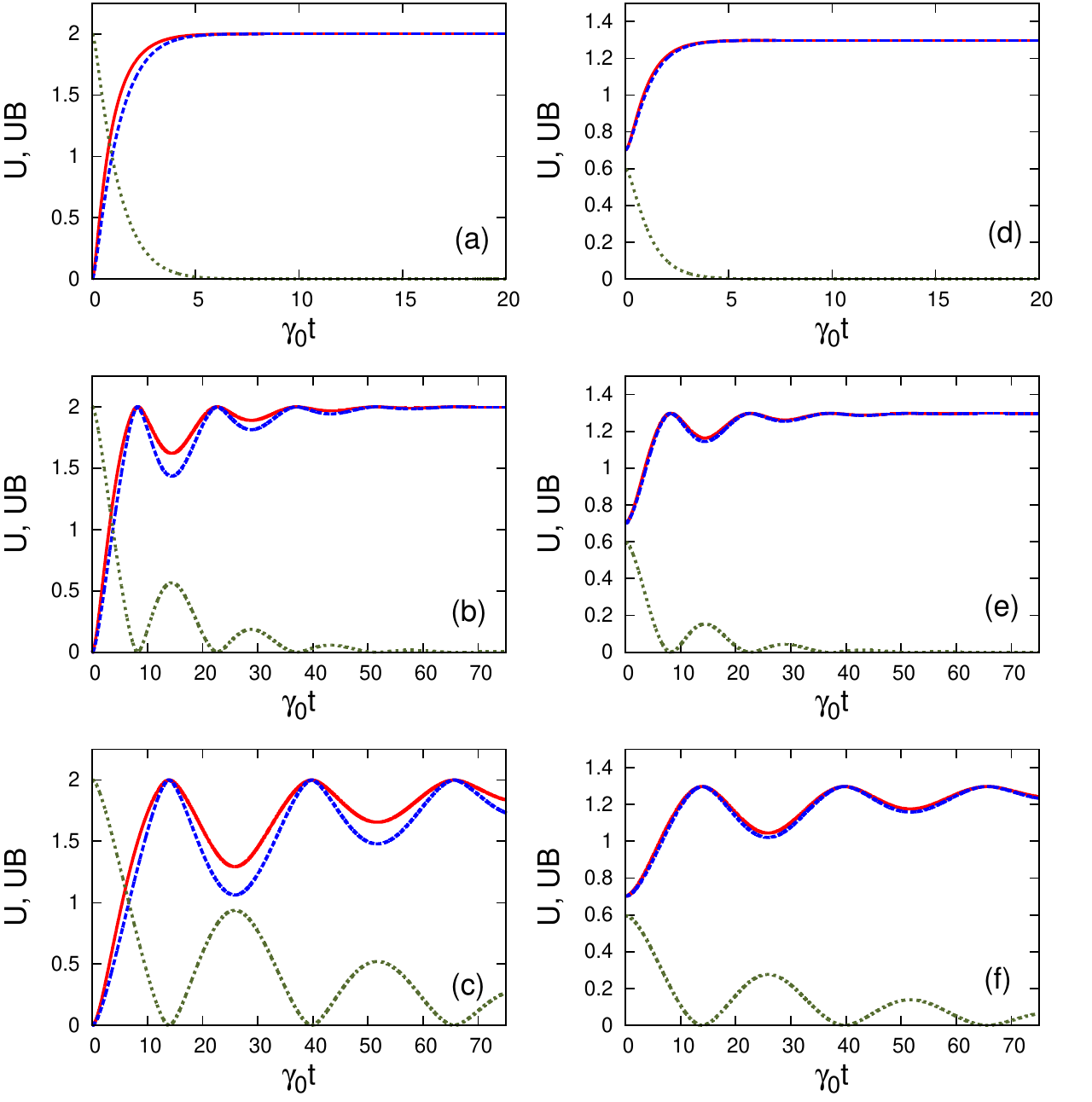}
\caption{Uncertainty $U(t)$ (red solid line), its lower bound $UB(t)$ (blue dashed line) and mutual information $I(\rho_{AB}(t))$ (green dotted line) versus scaled time $\gamma_0t$ when the memory particle $B$ is affected by zero temperature relaxation noise. We have $a=0.5$, $b=0$ and $c=0.5$ for the initial state of $AB$ in (a), (b) and (c), and $a=0.5$, $b=0.4$ and $c=0.1$ in (d), (e) and (f). The parameter controlling non-Markovianity is set as $\lambda/\gamma_0=3$ in (a) and (d), $\lambda/\gamma_0=0.1$ in (b) and (e), and $\lambda/\gamma_0=0.03$ in (c) and (f). Note that the plots in (a) and (d) shows the Markovian limit of the considered model.}
\label{fig3}
\end{figure}

Fig. \ref{fig3} displays our findings for the zero temperature relaxation noise on the memory particle $B$. In Fig. \ref{fig3}(a), (b) and (c), the initial state of the bipartite system $AB$ is taken as $a=0.5$, $b=0$ and $c=0.5$, which is the maximally entangled state. We set $\lambda/\gamma_0=3$ in (a) to show the Markovian limit of the model, and $\lambda/\gamma_0=0.1$ and $\lambda/\gamma_0=0.03$ in (e) and (f), respectively. We observe that, unlike in case of the dephasing model, both terms in $U(t)$ are non-zero here, and the maximally entangled state does not saturate the bound. On the other hand, we show the result of a similar analysis for the initial state $a=0.5$, $b=0.4$ and $c=0.1$ in Fig. \ref{fig3}(d), (e) and (f), where the uncertainty bound becomes almost tight. Our general conclusion about the memory effects reducing the lower bound of the memory-assisted entropic uncertainty relation can also be easily observed to hold here.

\textbf{Conclusion}-- In conclusion, considering a realistic setting where Bob's memory particle $B$ is an open system interacting with an external environment $E$, we have established a connection between the memory effects and the lower bound of Bob's uncertainty for two observables measured on Alice's system. We have demonstrated that memory effects, which have their roots in the non-Markovian features of the dynamical map, can be used to diminish the lower bound of the uncertainty relation. Furthermore, this reduction might in turn reduce Bob's ignorance about the outcomes of the measurements in $Q$ and $R$ bases on Alice's part, as demonstrated in two paradigmatic examples we present.

While specific results have been obtained on the effect of noise to the memory-assisted entropic uncertainty relation \cite{decoh}, we stress that our treatment provides a connection between the lower bound of uncertainty and the memory effects by means of an information theoretic definition of non-Markovianity. Thus, it is completely general and holds independently of any specific model for open quantum systems. In other words, the underlying relations between non-Markovian behavior and entropic uncertainty in the results obtained in Ref. \cite{decoh} can be consistently understood and unified through our model independent and information theoretic approach.

\textit{Acknowledgments}-- G. K. is grateful to Sao Paulo Research Foundation (FAPESP) for the BEPE postdoctoral fellowship given under grant number 2014/20941-7. S. M. and J. P. acknowledge financial support from the EU Collaborative project QuProCS (Grant Agreement 641277) and the Magnus Ehrnrooth Foundation.  J. P. also acknowledges the support from Jenny and Antti Wihuri Foundation.


\begin{thebibliography}{99}
\bibitem{hei27} HEISENBERG W., Z. Phys., \textbf{43} (1927) 173.
\bibitem{ken27} KENNARD E. H., Z. Phys., \textbf{44} (1927) 326.
\bibitem{robert29} ROBERTSON H. P., Phys. Rev., \textbf{34} (1929) 163.
\bibitem{deutsch83} DEUTSCH D., Phys. Rev. Lett., \textbf{50} (1983) 631.
\bibitem{entrobook} BIALYNICKI-BIRULA I. and RUDNICKI L., \textit{Statistical Complexity},
edited by K. D. Sen, (Springer, New York) 2011.
\bibitem{wehner10} WEHNER S. and WINTER A., New J. Phys., \textbf{12} (2010) 025009.
\bibitem{maasen88} MAASSEN H. and UFFINK J. B. M., Phys. Rev. Lett., \textbf{60} (1988) 1103.
\bibitem{berta10} BERTA M., CHRISTANDL M., COLBECK R., RENES J. M., RENNER R., Nature Phys., \textbf{6} (2010) 659.
\bibitem{exp1} PREVEDEL R., HAMEL D. R., COLBECK R., FISHER K., and RESCH K. J., Nature Phys., \textbf{7} (2011) 757.
\bibitem{exp2} LI C -F., XU J -S., XU X -Y., LI K., and GUO G. -C., Nature Phys., \textbf{7} (2011) 752.
\bibitem{openbook} BREUER H. -P. and PETRUCCIONE F., \textit{The Theory of Open Quantum Systems} (Oxford University Press, Oxford) 2007.
\bibitem{rivas10} RIVAS A., HUELGA S. F., PLENIO M. B., Phys. Rev. Lett., \textbf{105} (2010) 050403.
\bibitem{hou11} HOU S. C., YI X. X., YU S. X., and OH C. H., Phys. Rev. A, \textbf{83} (2011) 062115.
\bibitem{breuer09} BREUER H. -P., LAINE E.-M., PIILO J., Phys. Rev. Lett., \textbf{103} (2009) 210401.
\bibitem{lu10} LU X. -M., WANG X., and SUN C. P., Phys. Rev. A, \textbf{82} (2010) 042103.
\bibitem{luo12} LUO S., FU S., and SONG H., Phys. Rev. A, \textbf{86} (2012) 044101.
\bibitem{fanchini14} FANCHINI F. F., KARPAT G., \c{C}AKMAK B., CASTELANO L. K., AGUILAR G. H., FAR\'{\i}AS O. J., WALBORN S. P., SOUTO RIBERIO P. H., and de OLIVEIRA M. C., Phys. Rev. Lett., \textbf{112} (2014) 210402.
\bibitem{bylicka14} BYLICKA B., CHRU\'SCI\'NSKI D., MANISCALCO S., Sci. Rep., \textbf{4} (2014) 5720.
\bibitem{chr14} CHRU\'SCI\'NSKI D. and MANISCALCO S., Phys. Rev. Lett., \textbf{112} (2014) 120404.
\bibitem{vasile11} VASILE R., OLIVARES S., PARIS M. G. A., and MANISCALCO S., Phys. Rev. A, \textbf{83} (2011) 042321.
\bibitem{laine14}  LAINE E. -M., BREUER H. -P., and PIILO J., Sci. Rep., \textbf{4} (2014) 4620.
\bibitem{huelga12} HUELGA S. F., RIVAS A., and PLENIO M. B., Phys. Rev. Lett., \textbf{108} (2012) 160402.
\bibitem{chin12} CHIN A. W., HUELGA S. F., and PLENIO M. B., Phys. Rev. Lett., \textbf{109} (2012) 233601.
\bibitem{thorwart09} THORWART M. et al., Chem. Phys. Lett., \textbf{478} (2009) 234.
\bibitem{chin13} CHIN A. W., ROSENBACH J., CAYCEDO-SOLER F., HUELGA S. F., and PLENIO M. B., Nature Phys., \textbf{9} (2013) 113.
\bibitem{horodecki05} M. HORODECKI, J. OPPENHEIM, and A. WINTER, Nature, \textbf{436} (2005) 673.
\bibitem{notagree1} FANCHINI F. F., KARPAT G., CASTELANO L. K., and ROSSATTO D. Z., Phys. Rev. A, \textbf{88} (2013) 012105.
\bibitem{notagree2} ADDIS C., BYLICKA B., CHRU\'SCI\'NSKI D., and MANISCALCO S., Phys. Rev. A, \textbf{90} (2014) 052103.
\bibitem{haseli14} HASELI S., KARPAT G., SALIMI S., KHORASHAD A.S., FANCHINI F. F., \c{C}AKMAK B., AGUILAR G. H., WALBORN S. P., and SOUTO RIBEIRO P. H., Phys. Rev. A., \textbf{90} (2014) 052118.
\bibitem{daffer04} DAFFER S., WODKIEWICZ K., CRESSER J. D., MCIVER J. K., Phys. Rev. A, \textbf{70} (2004) 010304.
\bibitem{decoh} XU Z. Y., YANG W. L., and FENG M., Phys. Rev. A, \textbf{86} (2012) 012113; HU M. -L., and FAN H., Phys. Rev. A, \textbf{86} (2012) 032338; ZOU H. -M. et al., Int. J. Theor. Phys., \textbf{53} (2014) 4302; ZOU H. -M. et al., Phys. Scr., \textbf{89} (2014) 115101.
\end{thebibliography}
\end{document}